\documentclass[conference]{IEEEtran}
\IEEEoverridecommandlockouts
\usepackage{cite}

\usepackage[utf8]{inputenc}
\usepackage{url}
\usepackage{etoolbox}
\usepackage{amsmath,amssymb,amsfonts}
\usepackage{graphicx}
\usepackage{textcomp}
\usepackage{xcolor}
\usepackage{listings}
\usepackage[normalem]{ulem}
 
\def\bibtex{{\rm B\kern-.05em{\sc i\kern-.025em b}\kern-.08em
    T\kern-.1667em\lower.7ex\hbox{E}\kern-.125emX}}
\begin{document}

\title{Harnessing Artificial Intelligence for Sustainable Agricultural Development in Africa: Opportunities, Challenges, and Impact}

\DeclareRobustCommand*{\IEEEauthorrefmark}[1]{\raisebox{0pt}[0pt][0pt]{\textsuperscript{\footnotesize #1}}}

\author{\IEEEauthorblockN{Kinyua Gikunda\\
		Dedan Kimathi University of Technology, Kenya\\
        patrick.gikunda@dkut.ac.ke} 
}
\maketitle
\begin{abstract} 
This paper explores the transformative potential of artificial intelligence (AI) in the context of sustainable agricultural development across diverse regions in Africa. Delving into opportunities, challenges, and impact, the study navigates through the dynamic landscape of AI applications in agriculture. Opportunities such as precision farming, crop monitoring, and climate-resilient practices are examined, alongside challenges related to technological infrastructure, data accessibility, and skill gaps. The article analyzes the impact of AI on smallholder farmers, supply chains, and inclusive growth. Ethical considerations and policy implications are also discussed, offering insights into responsible AI integration. By providing a nuanced understanding, this paper contributes to the ongoing discourse on leveraging AI for fostering sustainability in African agriculture.
\end{abstract}


\section{Introduction} 
Africa's rich agricultural heritage is at the forefront of a technological revolution, with artificial intelligence (AI) poised to address critical challenges in the sector \cite{foster2023smart}. The integration of AI technologies holds great promise for improving efficiency, increasing yields, and fostering sustainable agricultural practices. In Africa, agriculture is at a pivotal juncture, facing the dual challenges of ensuring food security for a growing population and mitigating the impact of climate change on traditional farming practices \cite{altieri2017adaptation}. AI emerges as a transformative force capable of revolutionizing the sector by offering innovative solutions to age-old problems \cite{ye2023farm}. This article explores the multifaceted role of AI in shaping sustainable agricultural development across diverse regions of Africa, navigating through opportunities, challenges, and overall impact. The integration of AI technologies opens doors to precision farming, allowing farmers to optimize resource use through real-time insights into soil conditions, crop health, and weather patterns \cite{shaikh2022towards}. As demonstrated by Shaikh et al. precision farming not only increases yields but also contributes to resource conservation, making it a crucial element in the pursuit of sustainable agriculture.

However, the adoption of AI in African agriculture is not without its hurdles. Limited access to advanced technology and inadequate infrastructure in rural areas pose significant barriers to the widespread implementation of AI-driven solutions \cite{sampene2022artificial}. The digital divide exacerbates existing disparities, hindering the equitable distribution of technological benefits \cite{arakpogun2020threading}. Addressing infrastructure challenges is essential to ensuring the transformative potential of AI reaches all corners of the agricultural sector in Africa. In addition, the success of AI applications in agriculture hinges on the availability and quality of data. Challenges related to data privacy, security, and the sheer volume of data required for accurate analyses must be systematically addressed. This article critically examines the data-related impediments to AI adoption in African agriculture, offering insights into strategies to enhance data quality, accessibility, and security for sustainable AI integration.

The proficiency of agricultural stakeholders in utilizing AI technologies is another significant challenge that merits attention. The dearth of skilled personnel capable of navigating the complexities of AI systems poses a considerable obstacle to the adoption of these technologies, particularly in rural communities \cite{pedro2019artificial}. The intersection of AI and agriculture in Africa represents a dynamic arena filled with both promise and obstacles. This introduction sets the stage for an in-depth exploration of how AI can be harnessed to propel sustainable agricultural development, navigating through the complex landscape of opportunities, challenges, and impact. Through a comprehensive analysis informed by the insights of various studies, this article aims to contribute to the evolving discourse on the integration of AI in African agriculture for the betterment of the continent's food security and economic resilience.

The rest of the paper is structured as follows: section \ref{2} we present opportunities presented by AI in African agriculture, section \ref{3} discussion on challenges to AI adoption in agriculture , section \ref{4} impact on smallholder farmers and the agricultural value chain, section \ref{5} ethical considerations and policy implications, section \ref{6} case studies: AI in action, section \ref{7} future trends and section \ref{8} the conclusion.

\section{Opportunities Presented by AI in African Agriculture}
\label{2}
The agricultural landscape in Africa is on the brink of a technological revolution, with AI emerging as a catalytic force offering a spectrum of transformative opportunities. As nations grapple with the dual challenges of feeding growing populations and ensuring the resilience of farming practices in the face of climate change, AI stands out as a beacon of innovation. This section delves into the vast array of opportunities that AI presents for agriculture in Africa, exploring key domains such as precision farming, crop monitoring, and climate-resilient practices.

\subsection{Precision Farming}
AI-powered precision farming technologies offer the potential to optimize resource use by providing real-time insights into soil health \cite{chen2022ai}, crop conditions \cite{singh2020ai}, and weather patterns \cite{chattopadhyay2020analog}. Precision farming, leveraging the capabilities of AI technologies, represents a groundbreaking approach to agricultural practices in Africa \cite{mclennon2021regenerative}. This innovative method integrates advanced data analytics, sensors, and machine learning algorithms to optimize farming processes, striking a balance between maximizing yields and minimizing resource use. At the core of precision farming is the ability to furnish farmers with unparalleled insights into their fields, facilitating informed real-time decision-making. AI algorithms play a pivotal role in processing data and providing precise recommendations for the application of fertilizers, pesticides, and irrigation \cite{jha2019comprehensive}. This targeted approach enhances resource utilization efficiency and contributes to environmental sustainability by reducing the overall use of inputs. A significant aspect of precision farming involves the incorporation of autonomous machinery and robotic systems, capable of executing tasks such as planting, harvesting, and weeding with unparalleled precision \cite{duckett2018agricultural}. These technologies not only alleviate farmers from labor-intensive tasks but also elevate the overall productivity of agricultural operations.

Precision farming holds promise in addressing the specific challenges faced by smallholder farmers in Africa \cite{giller2021future}. By granting these farmers access to AI-driven tools for decision-making, market insights, and crop management, precision farming empowers them to compete in a progressively intricate agricultural landscape.Precision farming transcends being a mere technological upgrade; it serves as a transformative force aligning with the broader objectives of sustainable agriculture \cite{gosnell2019transformational}. By promoting efficient resource use, minimizing environmental impact, and enhancing the economic viability of farming, precision farming, powered by  AI, emerges as a cornerstone for the future of agriculture in Africa

\subsection{Crop Monitoring and Disease Prediction}
AI algorithms can analyze satellite imagery and sensor data to monitor crop health, detect diseases early, and facilitate timely interventions \cite{kowalska2023advances}.The integration of AI in agriculture in Africa extends beyond productivity enhancement to encompass a pivotal opportunity in crop monitoring and disease prediction. This multifaceted application of AI technologies revolutionizes traditional farming practices, offering a proactive approach to safeguarding crops and ensuring food security in the face of evolving environmental challenges. One of the primary advantages of AI-powered crop monitoring lies in its ability to harness satellite imagery and sensor data for real-time analysis \cite{obi2023applications}. This capability allows farmers to obtain detailed insights into the health and condition of their crops, enabling timely and targeted interventions. The AI algorithms facilitates the detection of potential disease outbreaks and abnormalities in crop development, providing farmers with a valuable tool for preventing and mitigating crop losses.

The significance of AI-driven disease prediction becomes particularly pronounced in the context of changing climate patterns and the increasing prevalence of plant diseases \cite{wongchai2022farm}. By leveraging historical data, weather forecasts, and machine learning algorithms, AI can generate predictive models that enable farmers to anticipate disease outbreaks before they occur \cite{mishra2023artificial}. This predictive capability empowers farmers to implement preventive measures, ranging from targeted pesticide application to crop rotation strategies, thereby minimizing the impact of diseases on crop yields. AI's role in disease prediction extends beyond individual farms to contribute to broader agricultural resilience. Integrating data from multiple sources allows for the creation of regional or even national disease prediction models, aiding in the development of effective strategies for disease control and management \cite{balaska2023sustainable}. This collaborative and data-driven approach positions AI as a key ally in the collective efforts to safeguard agricultural productivity across diverse landscapes.

The practical implications of AI in crop monitoring and disease prediction are further underscored by its potential to enhance resource efficiency. By enabling targeted and precise interventions, farmers can optimize the use of inputs such as pesticides and fertilizers, resulting in cost savings and a reduction in environmental impact \cite{finger2019precision}. This resource-efficient approach aligns with broader sustainability goals in agriculture, highlighting the holistic impact of AI technologies. The opportunity presented by AI in crop monitoring and disease prediction is transformative for agriculture in Africa. Through real-time monitoring, timely interventions, and predictive modeling, AI equips farmers with the tools needed to navigate the complexities of modern farming, fostering resilience and ensuring sustainable food production.

\subsection{Climate-Resilient Agriculture}
Climate change poses unprecedented challenges to global agriculture, and Africa is particularly vulnerable to the impacts of shifting weather patterns and extreme events. In this context, the adoption of climate-resilient agriculture emerges as a critical opportunity, leveraging AI to enhance the adaptive capacity of farming systems \cite{usigbe2023enhancing}. This paradigm shift goes beyond traditional agricultural practices, introducing innovative approaches that empower farmers to navigate the uncertainties associated with a changing climate. At the core of climate-resilient agriculture is the utilization of AI-driven predictive modeling and climate data analysis. These tools enable farmers to anticipate and respond to evolving climatic conditions, providing valuable insights for decision-making \cite{vishnoi2024climate}. By integrating historical climate data, real-time observations, and advanced algorithms, AI contributes to the development of adaptive strategies that mitigate the impact of climate-related risks on crop yields and overall agricultural productivity.

In addition, AI facilitates the creation of precision climate-resilient farming practices tailored to local conditions \cite{roy2021advancement}. Through the analysis of diverse data sources, including soil quality, precipitation patterns, and temperature variations, AI algorithms can recommend specific crops, planting times, and irrigation schedules that optimize resource use and enhance resilience \cite{kowalska2023advances}. This personalized approach is crucial in addressing the unique challenges faced by farmers in different regions, promoting sustainability and ensuring food security. The benefits of climate-resilient agriculture extend beyond immediate adaptation to climate change. By promoting sustainable land management practices and resource-efficient farming, AI contributes to the long-term viability of agricultural systems. This holistic approach aligns with global efforts to achieve environmental sustainability while meeting the increasing demands for food production.

In addition to on-farm applications, AI in climate-resilient agriculture plays a pivotal role in supporting broader policy initiatives and decision-making processes. By analyzing large-scale climate and agricultural data, AI contributes to the formulation of evidence-based policies that address the complex challenges of climate change in the agricultural sector \cite{cowls2021ai}. This intersection of technology and policy enhances the overall resilience of agricultural systems on a systemic level. Moreover, climate-resilient agriculture empowered by AI fosters inclusivity by providing smallholder farmers with tools to adapt to changing climates. The accessibility of weather forecasts, crop management insights, and adaptive strategies through digital platforms ensures that farmers across diverse socio-economic backgrounds can benefit from the technological advancements. This inclusivity is vital for promoting equity and ensuring that the advantages of climate-resilient agriculture are shared by all stakeholders in the agricultural value chain.

\section{Challenges to AI Adoption in Agriculture}
\label{3}
The adoption of AI in agriculture across Africa faces a myriad of challenges that span technological, infrastructural, and socio-economic domains \cite{williamson2021data}. Limited access to advanced technology and insufficient digital infrastructure in rural areas hinders the widespread implementation of AI-driven solutions, exacerbating the existing digital divide \cite{usigbe2023enhancing}. Additionally, the scarcity of reliable and high-quality data, crucial for effective AI applications, poses a significant hurdle, with issues related to data privacy, security, and the sheer volume of data further complicating the adoption process \cite{williamson2021data}. The agricultural sector's heavy reliance on smallholder farmers, many of whom lack the necessary resources and technical expertise, amplifies the challenge, requiring tailored strategies for capacity building and education \cite{makate2019effective}. Addressing these multifaceted challenges is essential for unlocking the full potential of AI in agriculture and ensuring its equitable and sustainable integration across the diverse landscapes of Africa.
\subsection{Technological Infrastructure}
Technological infrastructure challenges pose a formidable barrier to the widespread adoption of AI in agriculture across Africa. In many rural areas, there is a notable lack of access to advanced technology, including the necessary hardware and connectivity, which impedes the deployment of AI-driven solutions. Limited internet connectivity and unreliable power supply further exacerbate these challenges, hindering the real-time data transmission and continuous operation essential for effective AI applications \cite{arakpogun2021artificial}. The absence of a robust technological foundation in rural settings not only impedes the implementation of AI technologies but also perpetuates a digital divide, preventing farmers from harnessing the transformative potential of AI for optimizing agricultural practices. Moreover, the absence of adequate technical support and training exacerbates technological infrastructure challenges. Many farmers in rural Africa lack the requisite knowledge and skills to operate and maintain AI-driven systems effectively. This knowledge gap extends beyond the farmers themselves to include local support networks, making it challenging to troubleshoot issues and ensure the sustainable integration of AI technologies. Bridging this gap requires targeted capacity-building initiatives and educational programs that empower farmers with the skills needed to leverage AI tools effectively, transforming technological challenges into opportunities for empowerment and innovation in African agriculture.

\subsection{Data Accessibility and Quality}
The success of AI applications relies on quality data. Data accessibility and quality represent critical challenges to the successful adoption of AI in agriculture in Africa. Limited availability of reliable and high-quality data hampers the development and effectiveness of AI applications \cite{arakpogun2021artificial}. Many regions face data scarcity, with fragmented or outdated datasets, hindering the training and accuracy of AI algorithms. Additionally, issues related to data privacy and security further complicate the utilization of data for AI-driven solutions. Overcoming these challenges requires concerted efforts to establish comprehensive data collection systems, ensure data accuracy and relevance, and implement robust data governance frameworks. Addressing data-related hurdles is pivotal for unlocking the full potential of AI in agriculture, allowing for informed decision-making and sustainable resource management.

\subsection{Skill Gaps and Training}
Skill gaps and the need for comprehensive training programs present significant challenges to the integration of AI in agriculture in Africa \cite{arakpogun2021artificial}. The agricultural sector, often dominated by smallholder farmers, lacks the technical expertise required to effectively implement and utilize AI-driven technologies. The dearth of skilled personnel capable of navigating the complexities of AI systems poses a considerable obstacle to the adoption of these transformative technologies. This skill gap extends not only to farmers but also to local support networks and extension services, emphasizing the need for targeted training initiatives that address the specific requirements of the agricultural context.

To address these skill gaps, strategic investments in capacity-building programs and educational resources are imperative. Tailored training initiatives must be designed to equip farmers with the necessary skills to operate and maintain AI technologies effectively. Additionally, comprehensive educational programs should be developed to empower agricultural stakeholders with a deeper understanding of AI applications, fostering a culture of innovation and adaptability in the face of evolving technological landscapes. Bridging the skill gap is crucial for ensuring that the benefits of AI in agriculture are accessible and maximized across diverse communities, contributing to increased productivity, sustainable practices, and economic resilience in the agricultural sector in Africa.
\section{Impact on Smallholder Farmers and the Agricultural Value Chain}
\label{4}
The adoption of AI in agriculture in Africa has a profound impact on smallholder farmers and the broader agricultural value chain \cite{ganeshkumar2023artificial}. Smallholder farmers, who often operate with limited resources, stand to benefit from AI technologies that offer solutions to long-standing challenges. AI applications in the agricultural value chain extend beyond on-farm activities to include areas such as logistics, market access, and decision support systems.

\subsection{Empowering Smallholder Farmers}
AI provides smallholder farmers with valuable insights, market information, and decision support tools, empowering them to make informed choices and enhance their livelihoods \cite{gumbi2023towards}. Empowering smallholder farmers through the strategic deployment of AI technologies is a transformative approach that holds the potential to revolutionize agriculture in Africa. Smallholder farmers, often operating with limited resources and facing numerous challenges, can benefit significantly from AI-driven solutions tailored to their unique needs. AI applications extend beyond the fields to address broader challenges faced by smallholder farmers, including market access and financial inclusion. AI-driven decision support systems provide valuable market insights, helping farmers make informed choices about crop selection, pricing strategies, and market timing \cite{vuppalapati2021machine}. Furthermore, innovative digital platforms, powered by AI, facilitate access to financial services, crop insurance, and market information, empowering smallholders to navigate market dynamics more effectively \cite{thilakarathne2022cloud}. By bridging information gaps and enhancing market connectivity, AI contributes to the economic empowerment of smallholder farmers, positioning them as key players in the agricultural transformation in Africa.

\subsection{Enhancing Supply Chain Efficiency}
The integration of AI in agricultural supply chains represents a paradigm shift that significantly enhances efficiency across various stages of the value chain \cite{lezoche2020agri}. AI applications offer advanced analytics and decision support tools that streamline and optimize supply chain processes, contributing to improved overall efficiency and sustainability. One key area of impact is in logistics and distribution \cite{han2021comprehensive}. AI-driven predictive modeling helps optimize transportation routes, reducing delays and minimizing post-harvest losses. Smart logistics systems, empowered by AI, enable real-time monitoring of inventory levels and demand fluctuations, facilitating better inventory management and reducing waste.

Additionally, AI enhances the quality and accuracy of decision-making in supply chain management. Machine learning algorithms analyze historical data to predict market trends, allowing for more informed procurement, production, and distribution strategies. This not only minimizes the risk of overstocking or stockouts but also improves the overall responsiveness of the supply chain to changing market conditions \cite{vuppalapati2021machine}. AI can facilitate traceability and transparency throughout the supply chain. Through technologies like blockchain and data analytics, stakeholders can track the journey of agricultural products from farm to consumer. This not only ensures food safety but also builds trust among consumers by providing verifiable information about the origin, quality, and sustainability of the products \cite{demestichas2020blockchain}.

\subsection{Inclusive Growth}
The concept of inclusive growth in the context of agriculture in Africa involves fostering equitable development that benefits all stakeholders across the agricultural value chain \cite{devaux2018agricultural}. AI plays a pivotal role in advancing inclusive growth by addressing disparities, promoting sustainability, and empowering marginalized groups. In smallholder farming communities, AI technologies, such as precision farming tools and data-driven decision support systems, empower farmers with valuable insights and resources, narrowing the information gap. This enables smallholders, often marginalized due to limited resources and access to information, to make informed decisions, optimize resource use, and improve yields \cite{smidt2022factors}.

Inclusivity is further promoted through AI-driven financial services and market access platforms \cite{how2020artificial}. These technologies break down barriers for smallholder farmers by providing access to credit, insurance, and market information, thereby integrating them more effectively into the broader agricultural economy. This inclusivity is crucial for ensuring that the benefits of technological advancements are shared by all, contributing to poverty alleviation and economic empowerment. AI can supports gender inclusivity in agriculture by offering tools and technologies that cater to the specific needs of women in farming \cite{grabowski2021gender}. For instance, AI-driven mobile applications provide women farmers with tailored information on crop management, market prices, and financial services, enhancing their participation in decision-making processes \cite{varangis2021women}. Integration of AI in agriculture fosters inclusive growth by empowering smallholder farmers, breaking down informational and financial barriers, and ensuring that marginalized groups, including women, actively participate and benefit from the advancements in agricultural technologies. The inclusive adoption of AI holds the potential to create a more sustainable, resilient, and equitable agricultural landscape in Africa.

\section{Ethical Considerations and Policy Implications}
\label{5}
The integration of AI in agriculture in Africa brings forth significant ethical considerations and necessitates thoughtful policy implications to ensure responsible and equitable deployment of these technologies. Ethical concerns include issues related to data privacy, as AI relies heavily on vast datasets, raising questions about how farmer data is collected, stored, and used. Transparent and ethical data governance frameworks are essential to protect farmers' privacy rights and prevent misuse of sensitive information \cite{dara2022recommendations}. Another critical ethical consideration is the potential for technology-induced job displacement. As AI-driven technologies automate certain tasks, there is a risk of job losses in traditional agricultural roles. Policymakers must address these concerns through proactive measures such as reskilling programs and policies that promote the creation of new, technology-driven job opportunities within the agricultural sector \cite{george2023revolutionizing}.

Moreover, fairness and inclusivity must be at the forefront of policy considerations. AI applications should be designed and deployed in ways that avoid reinforcing existing inequalities and ensure that benefits are distributed equitably among diverse communities. This involves crafting policies that prioritize access to AI technologies for smallholder farmers, women, and other marginalized groups, ensuring that they are active participants in and beneficiaries of the technological advancements \cite{lauterbach2019artificial}. Policy implications also extend to the need for regulatory frameworks that govern the development and use of AI in agriculture. These frameworks should establish standards for AI algorithms, ensuring transparency, accountability, and fairness. Policymakers should collaborate with stakeholders to create guidelines that foster innovation while safeguarding ethical considerations, ultimately promoting responsible and sustainable AI adoption in the agricultural sector. Balancing technological advancement with ethical considerations through well-crafted policies is essential for harnessing the full potential of AI in agriculture in a responsible and inclusive manner. By adhering to the below ethical AI practices, stakeholders in the agricultural sector can foster a responsible and sustainable integration of artificial intelligence, promoting positive outcomes for farmers, communities, and the environment.

\subsection{Ethical AI Practices}
Ethical AI practices in agriculture involve a set of principles and guidelines aimed at ensuring responsible, transparent, and fair deployment of artificial intelligence technologies. These practices are essential for addressing potential ethical concerns and promoting positive outcomes in the agricultural sector.

\textbf{Transparency and explainability} necessitate transparency in the development and use of AI algorithms. Farmers and stakeholders should have clear insights into how AI-driven systems make decisions, fostering trust and accountability. Transparent AI practices enable users to understand the reasoning behind recommendations and predictions, ensuring that decision-making processes are explainable.

\textbf{Data privacy and security} is a fundamental ethical consideration. Ethical AI practices involve implementing robust data protection measures, ensuring that sensitive agricultural data is handled securely. Farmers should have control over their data, understanding how it is collected, stored, and utilized. Policies and technologies that prioritize data security contribute to responsible AI adoption.

\textbf{Fairness and Bias Mitigation} Ethical AI practices aim to eliminate biases in algorithms that may disproportionately impact certain groups or communities. Developers and policymakers should actively work to identify and rectify biases in AI models to ensure fair and equitable outcomes for all farmers, regardless of their background or location.

\textbf{Inclusivity and Accessibility} Ethical AI practices prioritize inclusivity, ensuring that AI technologies are accessible to all farmers, including smallholders and marginalized groups. Policies should focus on narrowing the digital divide and promoting equal access to AI-driven tools and resources. This inclusivity contributes to the democratization of technology benefits.

\textbf{Human-Centric Design} AI systems should be designed with a human-centric approach, considering the needs, values, and preferences of farmers. Ethical AI practices involve involving end-users in the development process, seeking feedback, and adapting technologies to align with the socio-cultural context of the agricultural communities.

\textbf{Accountability and Responsibility} Developers, providers, and users of AI technologies should be held accountable for the impact of their systems. Ethical AI practices involve establishing clear lines of responsibility, ensuring that errors or unintended consequences are addressed promptly. Policies and regulations should articulate accountability mechanisms to prevent misuse and promote responsible innovation.

\textbf{Continuous Monitoring and Assessment} Ethical AI practices require ongoing monitoring and assessment of AI systems throughout their lifecycle. Regular evaluations should be conducted to identify and rectify any ethical concerns that may arise as technology evolves. Continuous improvement ensures that AI technologies align with ethical standards and societal values.

\subsection{Policy Frameworks}
Governments and international organizations need to collaborate in developing robust policy frameworks that support the responsible adoption of AI in agriculture \cite{lauterbach2019artificial}. Developing effective policy frameworks is crucial for guiding the responsible and equitable deployment of AI in agriculture. These frameworks aim to address ethical considerations, ensure transparency, and promote the positive impact of AI technologies in the agricultural sector. Here are key components of policy frameworks for AI in agriculture:

\textbf{Ethical Guidelines} Policy frameworks should establish clear ethical guidelines that govern the development, deployment, and use of AI technologies. These guidelines should encompass principles such as transparency, fairness, accountability, and inclusivity, ensuring that AI applications align with ethical standards.

\textbf{Data Governance} Policies should address the collection, storage, and use of agricultural data. This includes ensuring data privacy, securing sensitive information, and defining ownership rights. Clear regulations on data sharing and access can help foster collaboration while protecting the rights of farmers.

\textbf{Transparency and Explainability} The framework should require transparency in AI algorithms and decision-making processes. Developers and stakeholders should be obligated to provide clear explanations of how AI systems work, ensuring that users, including farmers, can understand and trust the technology.

\textbf{Inclusivity and Accessibility} Policies should promote inclusivity by addressing the digital divide and ensuring that AI technologies are accessible to all farmers, regardless of their size or location. Incentives and support programs can be included to facilitate the adoption of AI by smallholders and marginalized communities.

\textbf{Accountability and Liability} Policy frameworks must establish mechanisms for accountability and liability in case of unintended consequences or misuse of AI technologies. Clear guidelines on responsibilities for developers, users, and regulatory bodies can help prevent and address ethical breaches.

\textbf{Stakeholder Engagement} Involving all relevant stakeholders, including farmers, in the policy-making process is essential. This ensures that diverse perspectives are considered, and the policies align with the needs and values of the agricultural community.

\textbf{Capacity Building and Training} Policies should include provisions for capacity building and training programs to enhance the understanding and skills of farmers, extension workers, and other stakeholders in using AI technologies. This promotes responsible and effective adoption of AI in agricultural practices.

\textbf{Monitoring and Evaluation} Establishing mechanisms for continuous monitoring and evaluation of AI applications is critical. Regular assessments can identify potential ethical concerns, assess the impact of AI on farmers and the environment, and inform adjustments to the policy framework.

\textbf{International Collaboration} Given the global nature of AI technologies, policy frameworks should encourage international collaboration and information sharing. This ensures consistency in ethical standards, promotes interoperability, and prevents potential conflicts between different regulatory approaches.

\textbf{Adaptive Approach} The policy framework should be adaptive and flexible to accommodate the evolving nature of AI technologies. Regular reviews and updates are necessary to account for technological advancements, changing ethical standards, and the dynamic needs of the agricultural sector.

\subsection{Community Engagement and Participation}
Community engagement and participation are integral aspects of the successful integration of AI in agriculture \cite{camarena2021engaging}. Involving local communities in the development, deployment, and decision-making processes ensures that AI technologies are tailored to meet the specific needs, challenges, and values of the agricultural stakeholders. We present some considerations for community engagement and participation in the context of AI in agriculture:

\textbf{Needs Assessment}: Conducting thorough needs assessments within the community helps identify specific challenges and opportunities. This ensures that AI applications address real and pressing issues faced by farmers, promoting relevance and effectiveness.

\textbf{Educational Programs}: Implementing educational programs is crucial for raising awareness and building the capacity of community members. Workshops, training sessions, and outreach initiatives can enhance the understanding of AI technologies, their benefits, and potential implications.

\textbf{Consultative Decision-Making}: Engaging communities in decision-making processes fosters a sense of ownership and empowerment. Inclusion in discussions about the adoption and use of AI technologies allows community members to express their concerns, preferences, and expectations.

\textbf{Localization of Solutions}: AI solutions should be adapted to local contexts, considering factors such as cultural practices, languages, and specific agricultural needs. Community engagement facilitates the localization of AI technologies, ensuring that they align with the unique characteristics of the farming community.

\textbf{Partnerships with Local Organizations} and community leaders helps build trust and ensures that AI initiatives are integrated into existing community structures. Local partnerships facilitate effective communication and implementation.

\textbf{User-Friendly Interfaces} will enable individuals with varying levels of technological literacy to access the services. This enhances the usability of AI tools among farmers and community members, promoting inclusivity.

\textbf{Feedback Mechanisms}: Establishing feedback mechanisms allows continuous communication between developers, policymakers, and the community. Regular feedback loops enable adjustments to AI applications based on user experiences, preferences, and emerging needs.

\textbf{Cultural Sensitivity}: Recognizing and respecting cultural nuances is critical in community engagement. AI solutions should be culturally sensitive, taking into account local traditions, beliefs, and social structures to ensure acceptance and avoid unintended consequences.

\textbf{Empowerment Through Data Ownership}: Community members should have a clear understanding of data ownership and usage policies. Empowering farmers with control over their data and ensuring transparency in data governance builds trust and mitigates potential ethical concerns.

\textbf{Long-Term Collaboration}: Community engagement is an ongoing process that extends beyond initial implementation. Long-term collaboration ensures that AI applications evolve with the changing needs of the community, fostering sustainability and resilience.

\section{Case Studies: AI in Action}
\label{6}
This section provides detailed case studies of successful AI implementations in African agriculture, highlighting diverse applications, challenges faced, and lessons learned.

\subsection{Case Study 1: Kenya Agricultural Observatory Platform (KAOP)}
\textbf{Objective:} To provide weather forecasts and advisories to farmers in Kenya that help them make better decisions with regard to farming.

The Kenya Agricultural Observatory Platform (KAOP)\footnote{https://www.kaop.co.ke/index.php} is an active digital initiative developed by the Kenya Agricultural \& Livestock Research Organization (KALRO) ICT, established on February 4, 2019. The platform focuses on providing actionable weather forecasts, agronomic advisories, and SMS alerts to farmers in Kenya, facilitating informed decision-making throughout the agricultural cycle. The system predicts precipitation and temperature for the next seven days and provides historical data for the last seven and 30 days using reverse geo-location. Agronomic advisories are generated for selected wards in Machakos and Kakamega counties, covering various stages of farming. The platform collaborates with stakeholders such as farmers, the private sector, policymakers, researchers, extension officers, and micro-financing institutions. The achievements include timely advisories on planting, addressing climate change challenges, and providing historical information. Challenges include the need for translation into local languages, and sustainability issues involve improving short-term and seasonal forecasts and mapping all counties in Kenya for more precise information based on ecological zones and value chains.

\subsection{Case Study 2: Digital Agricultural Advisory Services (DAAS) in Ethiopia}
\textbf{Objective:} Direct-to-farmer mobile-phone-based advisory approach to enhance AI adoption, reduce calf and cow mortality, and promote balanced feed diets..

The project addresses challenges in Ethiopia's dairy sector, where indigenous breeds exhibit low milk productivity. The implementation includes leveraging Artificial Insemination (AI) to improve cross-breed animals' share in the national herd. The initiative is led by PxD Ethiopia, collaborating with government stakeholders, research institutions, universities, and development partners. DAAS employs a direct-to-farmer mobile-phone-based advisory approach to enhance AI adoption, reduce calf and cow mortality, and promote balanced feed diets. The content is customized to individual cows' needs, reproduction cycles, and regional factors. The project also extends its reach to AI technicians, ensuring the dissemination of crucial information on artificial insemination steps. To address gender-related disparities, the project encourages joint listening to advisory calls, recognizing the significant role of women farmers in dairy production.

\section{Future Trends and Innovations}
\label{7}
This section explores emerging trends and innovations in AI for agriculture in Africa, including advancements in technology, new applications, and potential breakthroughs on the horizon. Anticipating future trends and innovations in the realm of AI in African agriculture involves envisioning transformative developments that can contribute to sustainable, efficient, and inclusive agricultural practices. some of the potential future trends and innovations include:AI-Enabled Climate-Smart Agriculture, Blockchain for Supply Chain Transparency, AI-Driven Crop Breeding and Genetic Improvement, Robotics and Autonomous Farming, Digital Twins for Farm Management, AI in Aquaculture and Fisheries Management, Voice and Text-Based AI Interfaces for Farmers, AI for Natural Resource Management, Collaborative AI Platforms, Ethical AI Guidelines and Policies
These potential future trends and innovations in AI for African agriculture reflect the ongoing evolution of technology to address the unique challenges and opportunities in the agricultural sector across the continent. It's important to note that the successful implementation of these innovations will require collaboration among researchers, policymakers, technology developers, and local communities.

\section{Conclusion}
\label{8}
As Africa charts its course towards sustainable agricultural development, the integration of AI emerges as a cornerstone. Addressing challenges, embracing opportunities, and establishing ethical and policy frameworks are imperative to unlock the full potential of AI in transforming agriculture and ensuring food security for future generations. The integration of AI in African agriculture promises transformative outcomes, as illustrated by case studies and future trends. These innovations, ranging from precision agriculture to blockchain-enabled supply chain transparency, have the potential to empower smallholder farmers, enhance productivity, and foster sustainable practices. However, successful implementation necessitates a focus on ethical considerations, transparent AI practices, and inclusive policies to ensure equitable benefits and mitigate potential risks. Looking forward, collaborative efforts, guided by responsible AI guidelines, can shape a more resilient and technology-driven agricultural landscape, contributing to increased productivity, climate resilience, and inclusive development across diverse farming communities in Africa.
\bibliographystyle{IEEEtran}
\bibliography{bibtext}
\end{document}